\documentclass[useAMS,usenatbib,referee]{mn2e}

\usepackage{epsf,graphicx,subfigure,lscape,longtable,txfonts,
}
\usepackage[paperwidth=595.27pt, paperheight=791.36pt, textwidth=470pt, textheight=670pt, left=63pt, top=65pt]{geometry}
\usepackage{rotating}
\usepackage{amssymb}
\usepackage[light]{}
\def\kms{\ifmmode{\rm km\,s}^{-1}\, \else km\,s$^{-1}$\,\fi}
\def\mujybm{$\rm{\mu}$Jy\,beam$^{-1}$\,}
\def\ltsim{\ifmmode\stackrel{<}{_{\sim}}\else$\stackrel{<}{_{\sim}}$\fi}
\def\gtsim{\ifmmode\stackrel{>}{_{\sim}}\else$\stackrel{>}{_{\sim}}$\fi}
\def\S4195{41.95+575}
\def\S4331{43.31+592}
\def\solmas{$\rm{M_\odot}$}
\def\solmasit{$M_\odot$}
\def\HII{{\sc {Hii}}}

\def\images{./finimages}

\title[Global VLBI and MERLIN combined monitoring of M82 SNR]{Wide-field Global VLBI and MERLIN combined monitoring of supernova remnants in M82.}
\author[D. Fenech, R. Beswick, T. W. B. Muxlow, A. Pedlar and M. K. Argo]{D. Fenech$^{1, 2}$ \thanks{E-mail:dmf@star.ucl.ac.uk} R. Beswick$^{2}$, T. W. B. Muxlow$^{2}$, A. Pedlar$^{2}$ and M. K. Argo$^{3}$\\
$^{1}$Department of Physics and astronomy, University College London\\
$^{2}$Jodrell Bank Centre for Astrophysics, School of Physics and Astronomy, Alan Turing Building, University of Manchester, M13 9PL\\
$^{3}$ICRAR, Curtin University of Technology, GPO Box UI987, Perth, Western Australia 6845, Australia}
\begin{document}

\date{}

\pagerange{\pageref{firstpage}--\pageref{lastpage}} \pubyear{}

\maketitle

\label{firstpage}

\begin{abstract}
{From a combination of MERLIN (Multi-Element Radio-Linked Interferometer Network) and global VLBI (Very Long Baseline Interferometry) observations of the starburst galaxy M82, images of 36 discrete sources at resolutions ranging from $\sim$3 to $\sim$80\,mas at 1.7\,GHz are presented. Of these 36 sources, 32 are identified as supernova remnants, 2 are {\HII} regions, and 3 remain unclassified. Sizes, flux densities and radio brightnesses are given for all of the detected sources. Additionally, global VLBI only data from this project are used to image four of the most compact radio sources. These data provide a fifth epoch of VLBI observations of these sources, covering a 19-yr time-line. In particular, the continued expansion of one of the youngest supernova remnants, 43.31+59.3 is discussed. The deceleration parameter is a power-law index used to represent the time evolution of the size of a supernova remnant. For the source 43.31+59.3, a lower limit to the deceleration parameter is calculated to be 0.53$\pm$0.06, based on a lower limit of the age of this source.} 
\end{abstract}


\section{Introduction}

Supernovae and supernova remnants (SNR) are a significant by-product of the intense star-formation found in starburst galaxies. As such, they play an important role in galaxy evolution and the feedback of material into the interstellar medium (ISM), and collectively via energy deposition into the ISM can drive massive superwinds as can be found in nearby galaxies such as M82 \citep{ohyama02}. Individual SNR are interesting in their own right and have been the subject of intense study for decades \citep[e.g.][]{bartbie03,marcaide09}. However, a population of SNR within a galaxy can provide information on the statistical properties of SNR evolution within the interstellar medium, as well as provide insight into the star-formation history of the galaxy via independent tracers of the supernova and star-formation rates.

Detailed investigations of SNR within the Galaxy are hindered both by the difficulty in determining accurate distances, and by the fact that all known Galactic SNR are over 150 years old \citep[][]{green08}. The use of radio interferometry allows high resolution observations of a starburst region and reveals sources that are otherwise obscured by the high levels of gas and dust associated with the star formation. In addition, such studies are free from selection effects as all of the SNR are at essentially the same distance and are observed with the same angular resolution and brightness sensitivity. 

For these reasons, a number of starburst galaxies have been well-studied using radio interferometric techniques, for example the southern hemisphere starburst galaxy, NGC\,253 \citep[e.g.][]{lenc06} and the more distant intense starburst of Arp\,220 \citep[e.g.][]{parra07}. However, as one of closest starburst galaxies, M82 \citep*[$\sim$3.2\,Mpc, ][]{burbidge64} presents a unique opportunity for a more detailed investigation of a young population of SNR. 

The first detailed radio observations of M82, began in the 1970s \citep{hargrave74, kronberg75} with the discovery of several compact sources within the central kpc. 
Subsequent observations (\citealp{Unger84}; Kronberg, Biermann \& Schwab 1985 \nocite{kronberg85}; \citealp{huang94}) have shown these to be part of a population of supernova remnants and {\HII} regions now approaching 100 identified sources. MERLIN observations of M82 have revealed the parsec-scale, shell-like structures of a large portion of these sources \citep[][]{muxlow94,fenech08} at 5\,GHz, though detailed observations at lower frequencies have been limited by the achievable resolutions and brightness temperature sensitivities of available interferometers. 

High angular resolution observations using the European VLBI Network (EVN), at a frequency of 1.7\,GHz began in 1997 utilising previous 1.4\,GHz observations from 1986, in order to observe and monitor the five most compact sources within M82. This began a now established global VLBI monitoring programme of these compact sources. The fifth epoch of this programme are presented here, extending the timeline of these observations to 19-yrs. 
However, these purely VLBI observations lack the surface brightness sensitivity to image all but the most compact objects. To remedy this in this work we present results from the first combined, simultaneous MERLIN and global VLBI observations of M82, which have enabled high surface brightness imaging at mas resolutions of the wider population of SNR and {\HII} regions.

We present the results of these combined observations in this paper. The detected population of sources is discussed in section \ref{pop}. The most compact sources are discussed individually as observed in the global VLBI only observations along with a detailed investigation of the possible deceleration of 43.31+59.2 in section \ref{expsec}. Section \ref{ism} discusses the supernova remnant environment in M82 and a summary is provided in section \ref{sum}.

\section{Observations and image processing}

This paper includes observations of M82 at 1.7\,GHz using both MERLIN and global VLBI as well as archival global VLBI and EVN observations from 1986, 1997, 1998 and 2001.

\subsection{New Observations}

Observations of M82 were performed with seven telescopes, of the MERLIN array on the 3rd Mar. 2005, at a frequency of 1.7\,GHz. Global VLBI observations were made simultaneously at the same frequency and included use of the ten telescopes of the VLBA (Very Long Baseline Array) in the USA as well as Westerbork, Medecina, Noto and Effelsberg antennas from the European VLBI Network (EVN). The global VLBI data also incorporated information from two MERLIN telescopes; the 76-m Lovell and 32-m Cambridge telescopes. The observations were made over a period of 18 hours with both the MERLIN and global VLBI arrays switching between M82 and the same phase calibration source J0958+65; this facilitates the combination of the two separate datasets.

\subsubsection{Global VLBI} The global VLBI data covers a range in frequency of 1.6595-1.6755\,GHz, split over 128 channels, each 250\,KHz in width. The data were correlated at JIVE (Joint Institute for VLBI in Europe) centred on the position of the most compact radio source 41.95+57.5. All subsequent data reduction and imaging was performed using the NRAO's {\sc {AIPS}} software. Amplitude calibration was performed using the system temperature and gain information provided for each telescope. The final delays, fringe rates and antenna gains were found for the calibration sources J0958+65 and J0927+39 and then applied to the target source, M82.

\begin{center}
\begin{table}
\caption{Observing information for the 1.7\,GHz observations made on Mar. 3 2005.}
\begin{center}
\begin{tabular}{|c|c|c|}
\hline
Array & MERLIN & Global VLBI \\
\hline
Central frequency (GHz) & 1.6646 & 1.6675\\
Total Bandwidth (MHz) & 8 & 16\\
Number of channels & 32 & 128\\
Bandwidth per channel (KHz) & 250 & 125\\
\hline
\end{tabular}
\label{obs}
\end{center}
\end{table}
\end{center}

\subsubsection{MERLIN} The MERLIN data were reduced using the MERLIN pipeline, utilising observations of the calibration source, 3C84 to set the flux density scale and the point source calibrator J0927+39 to perform a bandpass calibration. These MERLIN observations were reduced using standard routines to apply phase reference solutions from the phase calibrator J0958+65 to M82. Both the global VLBI and MERLIN observations were made using spectral channels to facilitate wide-field imaging, the channel setup used in each case is described in Table \ref{obs}.

\subsubsection{Combination} Following the separate calibration some preparation was required to enable combination. This included the splitting of the two datasets into their individual spectral channels which were then recombined to ensure the frequencies of the datasets matched (within 0.1\%), whilst not averaging the channels to enable wide-field imaging with the global VLBI data. This process avoids introducing the effects of bandwidth smearing. The flux density scales were then checked using th LT-CM baseline, common to both the MERLIN and global VLBI data, for observations of the calibration sources. A small positional correction of the global VLBI data was performed to ensure coincidence with the reference position of the MERLIN data. Following this correction, the two datasets were coincident to within $\lesssim$ 0.10\,mas. The datasets were then combined using the {\sc{aips}} task {\sc{dbcon}}.

\begin{figure}
\begin{center}
\includegraphics[width=7cm]{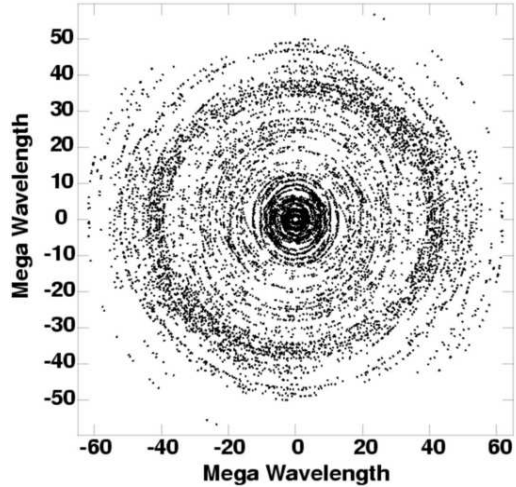}
\caption{The spatial frequency coverage of the combined MERLIN and global VLBI observations at 1.7\,GHz. The plot shows Every 500th visibility following the splitting of the data into individual spectral channels and subsequent combination.} 
\label{combuv}
\end{center}
\end{figure}

\subsection{Archive VLBI and EVN observations}

Four previously published datasets have been included in the analysis presented here. EVN observations from 1986 and 1997 initially published in \cite{pedlar99} and global VLBI observations from 1998 and 2001 published in \cite{mcdonald01} and \cite{beswick06}. The telescopes included in each of these observations are listed in Table \ref{globs} as well as those from the most recent epoch. Full details of the reduction procedures used for each of these datasets is described in the literature. The 1997, 1998 and 2001 observations all utilised very similar techniques for the observational and data reduction processes. The 1986 observations differed slightly in that the observing frequency used was 1.4\,GHz and the flux density scale was based on the total flux density of 41.95+57.5, at the epoch of these observations.

\begin{table}
\begin{center}
\caption{A summary of the EVN and global VLBI observations of M82.}
\begin{tabular}{|c|c|c|c|c|}
\hline
Epoch & Date & Array & Frequency & Reference\\
\hline
1 &11 Dec. 1986 (1986.95) & Ef, Jb, Wb, Mc & 1.4\,GHz& \cite{pedlar99}\\
2& 02 Jun. 1997 (1997.42) & Ef, Jb, Mc, Nt, On, Wb, Cm, Tr& 1.7\,GHz &\cite{pedlar99}\\
3 & 28 Nov. 1998 (1998.91) & VLBA, Y1, Ro, Go, Ef, Jb, Mc, Nt, On, Wb, Tr& 1.7\,GHz & \cite{mcdonald01}\\
4 & 23 Feb. 2001 (2001.15) & VLBA, Y1, Ro, Ef, Jb, Mc, Nt, On, Wb, Tr& 1.7\,GHz & \cite{beswick06}\\
5 & 03 Mar. 2005 (2005.17) & VLBA, Jb, Ef, Cm, Wb, Nt, Mc + MERLIN & 1.7\, GHz & This paper\\
\hline
\end{tabular}
\label{globs}
\end{center}
\end{table}

\subsection{Imaging}

Following combination, the MERLIN+global data were imaged using the {\sc {aips}} task {\sc {imagr}} and deconvolved with the H\"{o}gbom cleaning algorithm \citep{hogbom74}. In order to image each source in the most suitable way, a number of resolutions were used. Initially, a beam size of 50\,mas was used together with a cellsize of 10\,mas and a robustness weighting of 0. The whole of the central 700\,pc of M82 was imaged in this way using ten 1024$\times$1024 fields. These 50\,mas field images have a 1 ${\rm\sigma}$ rms noise level of $\sim$310\,\mujybm and a brightness temperature sensitivity level of $\sim$5900\,K.
Images with resolutions between 25 and 50\,mas (in 5\,mas steps) as well as $60,\,70$ and 80\,mas using a Gaussian taper to give the required resolution were also produced. 
The final parameters used to produce the images presented in section \ref{Im} are listed in Table \ref{vals}.

The observational setup for this experiment has been chosen to provide a well-sampled, evenly spaced spatial frequency coverage with minimal gaps (see Fig. \ref{combuv}), achieved in this case using a large number of telescopes to perform a full imaging track of a circumpolar source. This limits any image fidelity or brightness sensitivity issues as a result of an incomplete aperture coverage and represents the state-of-the-art in long baseline array observations. Multiple resolutions were used to provide the best image of each source and ensure that the Gaussian tapering applied did not result in any significant change in the observed source structures over the resolutions sampled.

\begin{landscape}
\begin{figure}
\begin{center}
\includegraphics[width=12cm,angle=270]{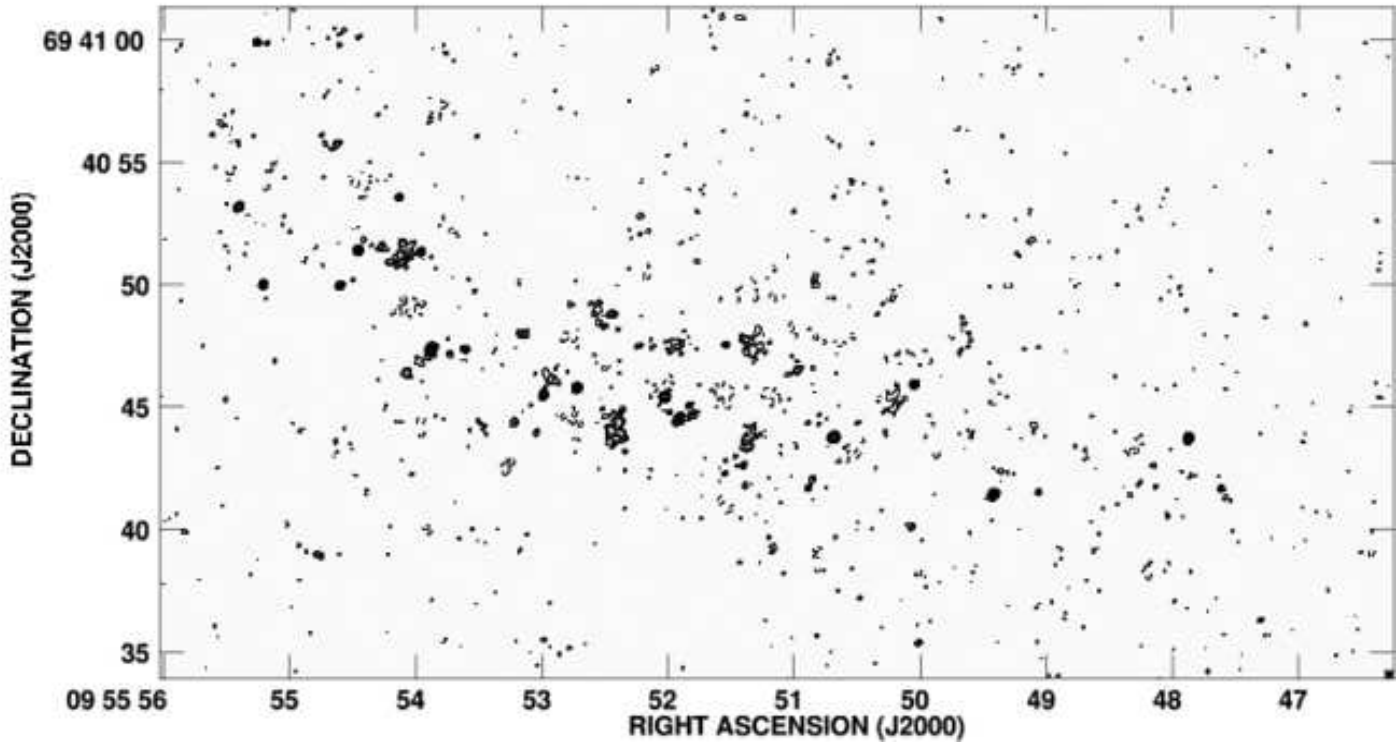}
\caption{A MERLIN 1.7\,GHz contour image of M82 from the observations presented here, restored with a 130\,mas beam,which is plotted in the bottom right-hand corner. The contours are plotted at $-1,1,2,4,6,8,10,12,14,16,18,20,25,30,35,40\,\times\,250$\,\mujybm.} 
\label{18cm-mer}
\end{center}
\end{figure}
\end{landscape}

A MERLIN only image has been used to measure the flux density values presented in Table \ref{VMtab}. This naturally weighted 4096$\times$2048 image has a restoring beam of 130\,mas and a cellsize of 30\,mas, a contour image of which is shown in Fig. \ref{18cm-mer}. The rms noise over source free areas of this MERLIN-only image is $\sim$55\,\mujybm.

The global VLBI (not including MERLIN) datasets were imaged using {\sc{imagr}}, applying an appropriate taper, to produce 512$\times$512 images, centred on 41.95+57.5, 43.18+58.3, 43.31+59.2, 44.01+59.6 and 45.17+61.2, restored with a 15\,mas circular beam. This enabled effective comparison with the original EVN dataset images with a matched angular resolution. Images were also produced using only the global VLBI epochs, with a cellsize of 1\,mas and a robustness weighting of 0 at suitable resolutions for each source. The source 41.95+57.5 was imaged using a circular 3.3\,mas beam and 43.31+59.2 was imaged with a circular beamsize of 4\,mas. The remaining three sources (43.18+58.3, 44.01+59.6 and 45.17+61.2) were imaged using resolutions of 8 and 10\,mas. 

\section{Source structures at 1.7\,GHz}\label{Im}

Contour and grey-scale images of the individual sources made from the combined MERLIN+global VLBI data are presented in Fig. \ref{conts}. Several shell or partial-shell structures can be clearly identified. The global VLBI and EVN images of the most compact sources can be found in section \ref{compact}, where the sources are discussed individually.

\begin{table}
\begin{center}
\caption[The beam size and corresponding noise levels of the MERLIN and global VLBI combined images used to produce the contour plots of the individual sources at 1.7\,GHz.]{The beam size and corresponding noise levels of the the MERLIN and global VLBI combined images used to produce the contour plots of the individual sources at 1.7\,GHz.}
\begin{tabular}{|c|c|}
\hline
Resolution (mas) & rms noise level (\mujybm)\\
\hline
25 & 27.0 \\
30 & 28.0 \\
35 & 29.6 \\
40 & 32.5 \\
50 & 31.1 \\
\hline
\end{tabular}
\label{vals}
\end{center}
\end{table}

Sources have been included with peak flux density of $\geq 5\sigma$ in both the MERLIN only image and the combined 50\,mas resolution images, which corresponds to $\geq275\,$\mujybm and $\geq$150\,\mujybm\, respectively, providing a sample of 36 sources. 

The source sizes have been measured in accordance with the observed source structure. The single peaked sources have been measured using Gaussian fitting to find the full width half maximum. The obvious shell-structured sources with a peak flux density (in the combined images) of $>$300\,\mujybm have been measured using radial profiles averaged in azimuth and measured to 50\% of the peak. The weaker more extended sources (peak flux density $>$ 250\,\mujybm) have been measured using the flux density profiles produced by drawing slices across the source. The values quoted are average diameters or a major by minor axis, wherever appropriate. An example of the measured sizes of two sources using all of the described methods is shown in Table \ref{meths} . This illustrates that the methods used provide approximately equivalent measures of the source size, whilst utilising the most appropriate procedures for a given source structure. 

The contour images in Fig. \ref{conts} are centred on the Right Ascension and Declination position listed in Table \ref{VMtab} as measured from the combined data. The flux density measurements and the deconvolved sizes for each source are also presented in this table. The integrated flux density was measured over an area matching the 3$\sigma$ size of the source in the MERLIN only image and a suitable subtraction made to account for the background emission in the vicinity of the source.

\begin{table}
\begin{center}
\caption[An example of two sources whose diameters measured using all methods for comparison.]{An example of two sources whose diameters have been measured using all methods for comparison. The errors, included in parentheses, represent those quoted from the fitting procedure, and those estimated from the measurements of the radial profile.}
\begin{tabular}{|c|c|c|c|c|}
\hline
\multicolumn{1}{c}{Source} &
\multicolumn{1}{c}{Resolution} &
\multicolumn{3}{c}{Deconvolved Diameter (mas)} \\
 & (mas) & Gaussian Fit (average) & Radial profile & Flux density slices\\
\hline
\hline
44.01+59.6 & 50 & 50.8(0.3) & 47(2) & 50\\
41.30+59.6 & 50 & 78(2) & 77(2) & 75\\
\hline
\end{tabular}
\label{meths}
\end{center}
\end{table}

\clearpage
\begin{tabfonta}
\begin{figure}
\begin{center}
\includegraphics[angle=0,width=17.0cm]{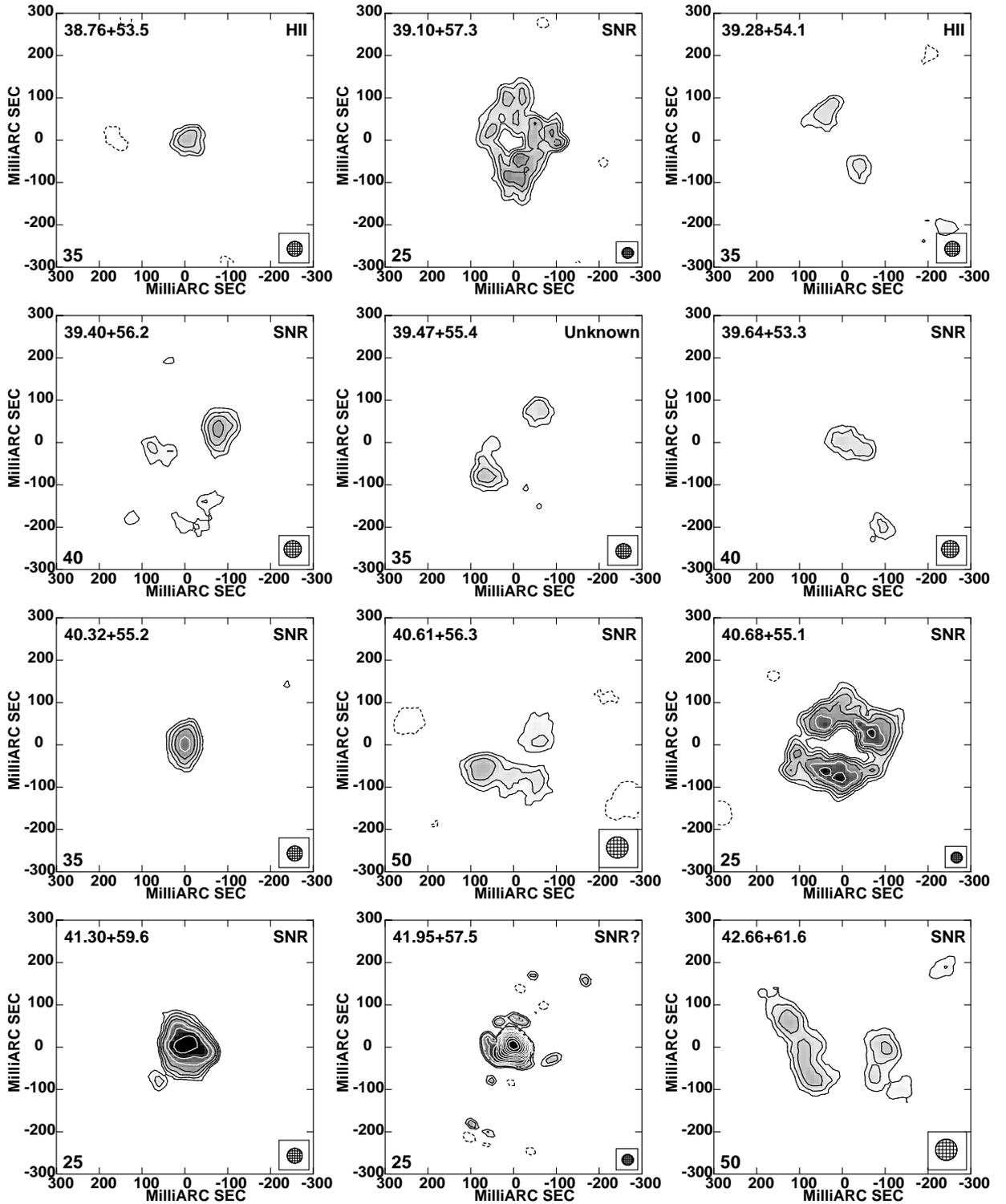}

\caption{Grey-scale and contour images of the individual sources produced from the combined MERLIN+global VLBI data. The grey-scale ranges from $130-600$\,\mujybm\, and the contours are plotted at $-1,\,1,\,1.414,\,2,\,2.828,\,4,\,5.656,\,8,\,11.282\,\times\, 3\rm{\sigma}$ noise level (see Table \ref{vals}). The beam size is listed numerically in units of milliarcseconds in the bottom left-hand corner and graphically in the bottom right-hand corner of each image. The source 41.95+57.5 has a grey-scale ranging from $200-800$\,\mujybm\, and contour levels plotted at multiples of 0.2\,mJy$\rm{bm^{-1}}$. }
\label{conts} 
\end{center}
\end{figure}
\end{tabfonta}

\begin{figure}
\begin{center}
\includegraphics[angle=0,width=15.5cm]{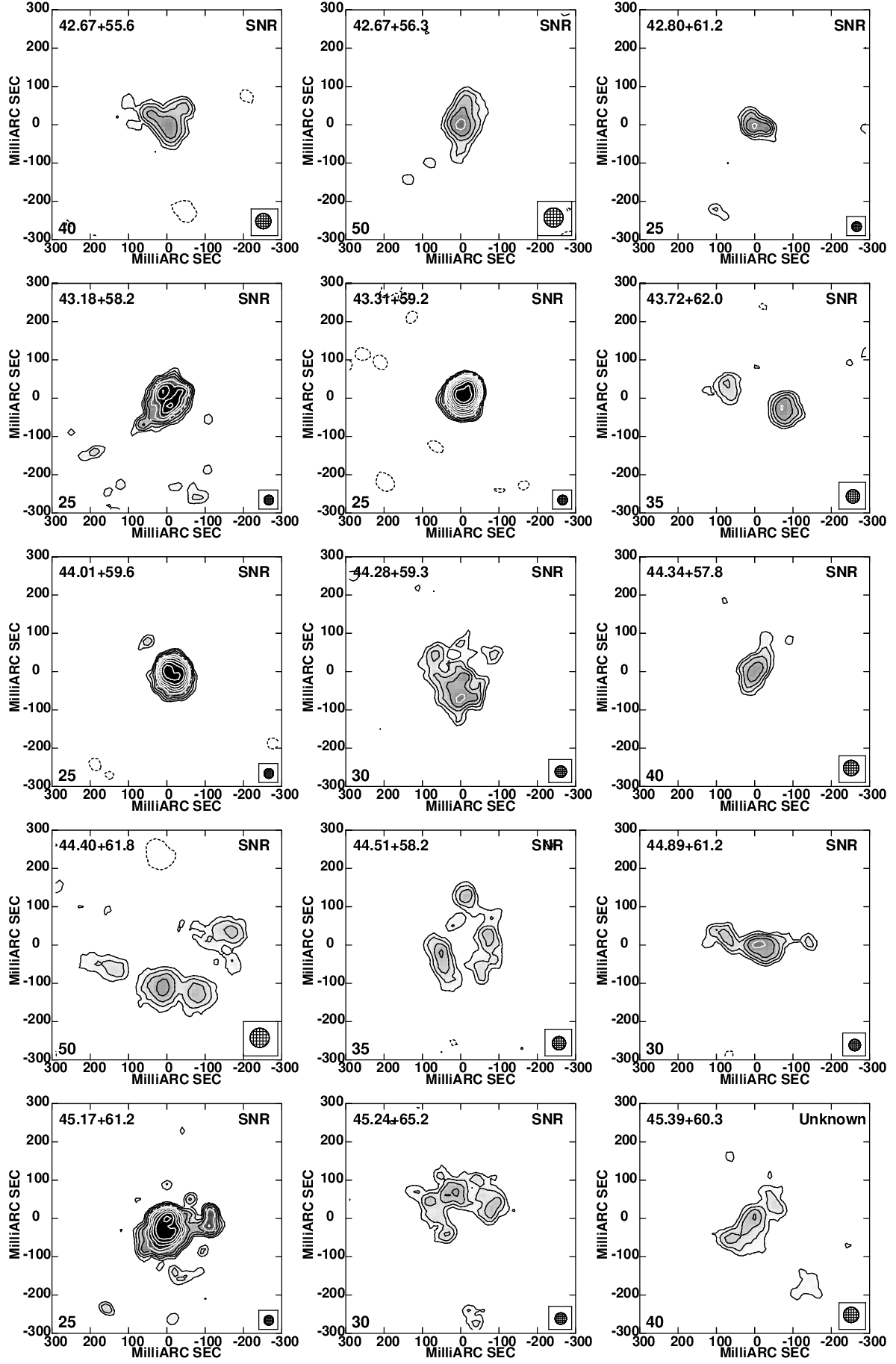}
\it{\\Figure 3 continued}
\end{center}
\end{figure}

\begin{figure}
\begin{center}
\includegraphics[angle=0,width=17.0cm]{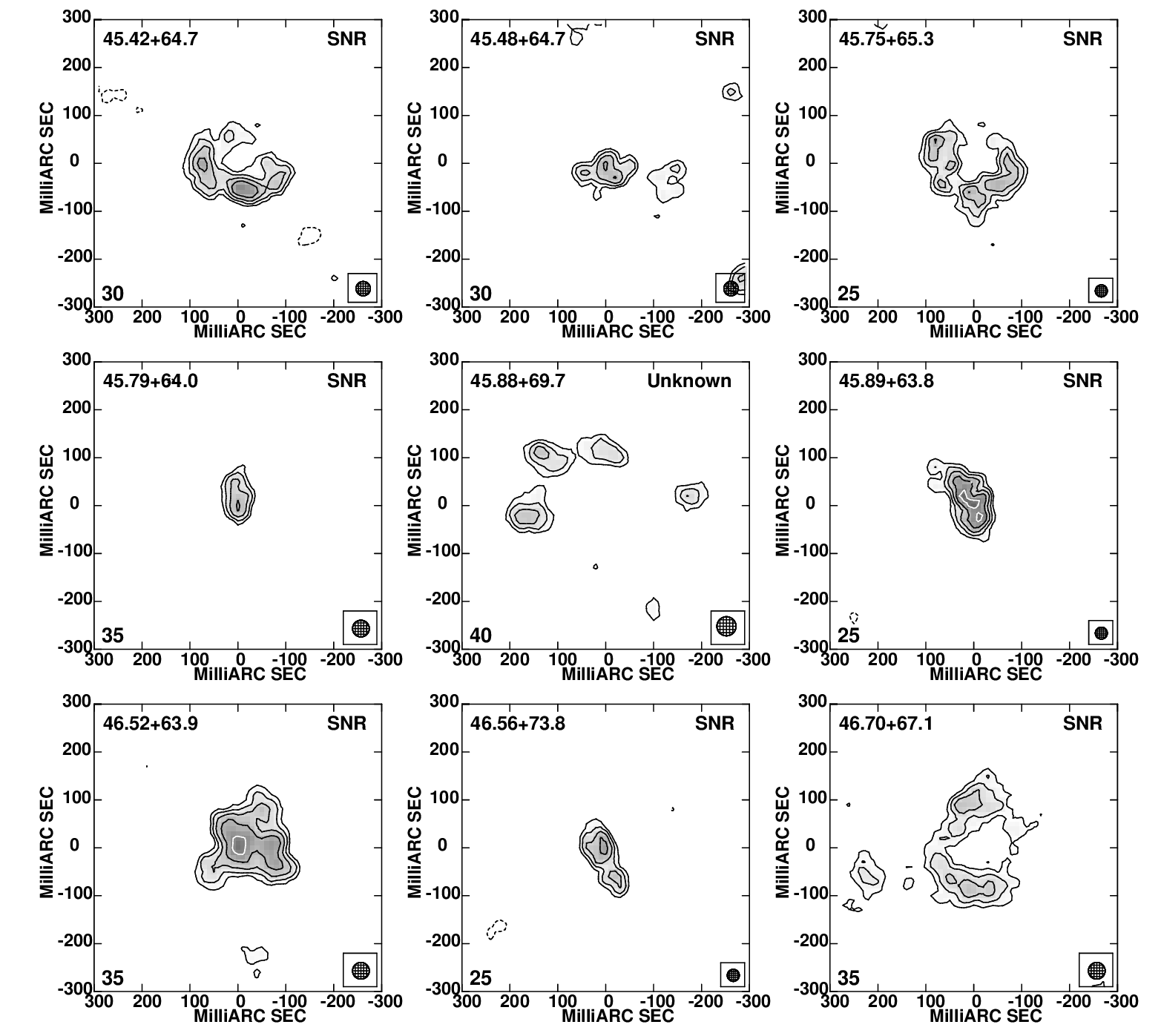}
\it {Figure 3 continued}
\end{center}
\end{figure}

\clearpage
\begin{onecolumn}
\renewcommand{\thefootnote}{\fnsymbol{footnote}}

\begin{tabfonta}
\begin{center}
\begin{longtable}[h]{lrrcccc}
\caption[]{Flux densities and deconvolved sizes of the sources detected in the global VLBI+MERLIN dataset. The names of each source are a combination of their B1950 Right Ascension seconds and Declination arcseconds, offset from $\rm{09^{h}\,51^{m}\,00^{s}}$ and $+69^{o}\,54'\,00''$, following the convention of \cite{kronberg85}. The J2000 Right Ascension and Declination are as measured in the combined data and are offset from  $\rm{09^{h}\,55^{m}\,00^{s}}$ and $+69^{o}\,40'\,00''$. The flux densities are measured from the MERLIN-only data. Errors are shown in brackets where appropriate.}
\label{VMtab}
\tabularnewline 
\hline
  \multicolumn{1}{c}{Name} &
  \multicolumn{1}{c}{RA} &
  \multicolumn{1}{c}{Dec} &
  \multicolumn{2}{c}{Diameter} &
  \multicolumn{1}{c}{Peak Flux Density} &
  \multicolumn{1}{c}{Total Flux Density} \\
& J2000 & J2000 & mas & pc & (mJy/beam) & (mJy) \\
\hline
\hline
\endfirsthead

\multicolumn{7}{l}
{{\it \tablename\ \thetable{} -- continued from previous page}}\\

\hline
  \multicolumn{1}{c}{Name} &
  \multicolumn{1}{c}{RA} &
  \multicolumn{1}{c}{Dec} &
  \multicolumn{2}{c}{Diameter} &
  \multicolumn{1}{c}{Peak Flux Density} &
  \multicolumn{1}{c}{Total Flux Density}\\
& J2000 & J2000 & mas & pc & (mJy/beam) & (mJy)\\
\hline	
\hline 
\endhead
\multicolumn{7}{l}{{\it Continued on next page}} \endfoot
\endlastfoot
\hline

38.76+53.5 & 47.53 & 39.94 & 69(9)$\times$46(7) & 1.1$\times$0.7 & 0.36\,(0.03) & 0.52\,(0.08) \\
39.10+57.3 & 47.88 & 43.72 & 229 & 3.5 & 3.63\,(0.05) & 5.74\,(0.10)\\
39.28+54.1 & 48.04 & 40.59 & 164 & 2.5 & 0.27\,(0.06) & 0.40\,(0.11)\\
39.40+56.2 & 48.16 & 43.00 & 197$\times$63 & 3.1$\times$1.0 & 0.36\,(0.05) & 0.65\,(0.09)\\
39.47+55.4 & 48.26 & 41.91 & 251$\times$53 & 3.9$\times$0.8 & 0.29\,(0.07) & 0.56\,(0.13)\\
39.64+53.3 & 48.40 & 39.82 & 177$\times$93 & 2.7$\times$1.4 & 0.27\,(0.03) & 0.41\,(0.05)\\
40.32+55.2 & 49.06 & 41.53 & 81(7)$\times$50(5) & 1.3$\times$0.8 & 0.79\,(0.06) & 0.95\,(0.11)\\
40.61+56.3 & 49.37 & 42.44 & 250 & 3.9 & 0.39\,(0.06) & 0.64\,(0.09)\\
40.68+55.1 & 49.42 & 41.43 & 182 & 2.8 & 6.48\,(0.06) & 11.33\,(0.17)\\
41.30+59.6 & 50.05 & 45.92 & 65 & 1.0 & 3.39\,(0.06) & 3.64\,(0.11)\\
41.95+57.5 & 50.69 & 43.76 & 28.5(0.1)$\times$21.7(0.1) & 0.4$\times$0.3 & 36.19\,(0.08) & 38.25\,(0.21)\\
42.66+51.6 & 51.39 & 47.80 & 207 & 3.2 & 0.93\,(0.04) & 1.83\,(0.08)\\
42.67+55.6 & 51.39 & 41.80 & 99 & 1.5 & 0.73\,(0.05) & 0.97\,(0.09)\\
42.67+56.3 & 51.40 & 42.63 & 132(13)$\times$61(7) & 2.0$\times$0.9 & 0.92\,(0.04) & 1.08\,(0.07)\\
42.80+61.2 & 51.55 & 47.54 & 71$\times$40 & 1.1$\times$0.6 & 0.87\,(0.05) & 0.90\,(0.06)\\
43.18+58.2 & 51.91 & 44.57 & 107 & 1.7 & 6.90\,(0.07) & 8.39\,(0.14)\\
43.31+59.2 & 52.03 & 45.42 & 53.6(0.3)$\times$44.1(0.2) & 0.8$\times$0.7 & 22.66\,(0.16) & 23.55\,(0.35)\\
43.72+62.0 & 52.45 & 48.80 & 208$\times$56 & 3.2$\times$0.9 & 1.08\,(0.05) & 1.65\,(0.09)\\
44.01+59.6 & 52.73 & 45.78 & 56.7(0.6)$\times$46.6(0.5) & 0.9$\times$0.7 & 11.43\,(0.06) & 12.32\,(0.12)\\
44.28+59.3 & 52.99 & 45.49 & 156 & 2.4 & 2.12\,(0.05) & 2.84\,(0.09)\\
44.34+57.8 & 53.05 & 43.94 & 118(13)$\times$51(6) & 1.8$\times$0.8 & 0.65\,(0.06) & 0.66\,(0.10)\\
44.40+61.8 & 53.15 & 48.04 & 235 & 3.6 & 1.14\,(0.06) & 2.25\,(0.13)\\
44.51+58.2 & 53.23 & 44.36 & 234 & 3.6 & 0.95\,(0.05) & 1.53\,(0.10)\\
44.89+61.2 & 53.61 & 47.36 & 223$\times$54 & 3.5$\times$0.8 & 1.24\,(0.05) & 1.97\,(0.11)\\
45.17+61.2 & 53.88 & 47.43 & 74$\times$56 & 1.1$\times$0.9 & 14.23\,(0.07) & 17.60\,(0.16)\\
45.24+65.2 & 53.96 & 51.29 & 223 & 3.4 & 1.31\,(0.06) & 1.59\,(0.10)\\
45.39+60.3 & 54.07 & 46.50 & 172$\times$49 & 2.7$\times$0.8 & 0.88\,(0.06) & 1.73\,(0.13)\\
45.42+64.7 & 54.13 & 53.58 & 201 & 3.1 & 1.24\,(0.05) & 1.54\,(0.09)\\
45.48+64.7 & 54.22 & 50.94 & 116$\times$55 & 1.8$\times$0.9 & 0.92\,(0.06) & 1.04\,(0.08)\\
45.75+65.3 & 54.46 & 51.43 & 222(10) & 3.4 & 2.30\,(0.08) & 3.35\,(0.16)\\
45.79+64.0 & 54.50 & 50.20 & 100(13)$\times$47(8) & 1.6$\times$0.7 & 0.27\,(0.03) & 0.31\,(0.05)\\
45.88+69.7 & 54.60 & 55.75 & 405$\times$183 & 6.3$\times$2.8 & 0.57\,(0.07) & 1.08\,(0.14)\\
45.89+63.9 & 54.60 & 49.98 & 118$\times$30 & 1.8$\times$0.5 & 2.12\,(0.01) & 2.37\,(0.01)\\
46.52+63.9 & 55.22 & 50.01 & 172$\times$139 & 2.7$\times$2.2 & 2.00\,(0.05) & 2.60\,(0.09)\\
46.56+73.8 & 55.26 & 59.91 & 135$\times$47 & 2.1$\times$0.7 & 1.26\,(0.07) & 1.40\,(0.12)\\
46.70+67.1 & 55.41 & 53.17 & 260 & 4.0 & 1.22\,(0.06) & 2.18\,(0.14)\\

\hline
\end{longtable}
\end{center}
\end{tabfonta}
\end{onecolumn}

\subsection{The source population}\label{pop}

\begin{figure}
\begin{center}
\includegraphics[width=12cm]{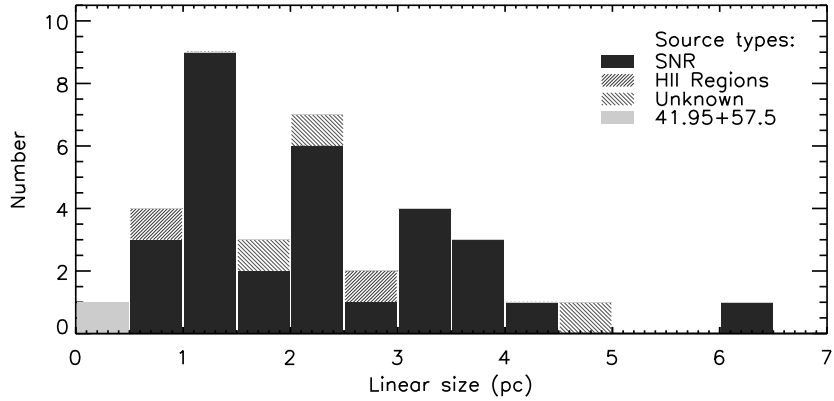}
\caption[A histogram showing the observed diameters of the SNR, {\HII} regions and the unknown sources at 1.7\,GHz.]{A histogram showing the observed diameters of the SNR, {\HII} regions and the unknown sources at 1.7\,GHz.}
\label{hist18cm}
\end{center}
\end{figure}

A total of 36 sources have been detected in these combined observations, 31 of which are previously identified supernova remnants with diameters ranging from 0.4 to 4.0\,pc and a mean size of 2.1\,pc \citep[identifications taken from catalogues in][]{wills97,allen99,mcdonald02,rodr04}. The remaining five sources consist of two {\HII} regions and three as yet unidentified sources. The two detected {\HII} regions are 38.76+53.5 and 39.28+54.1, originally identified by \cite{mcdonald02} with brightness temperatures $\sim$1100\,K. These are shown in the histogram of source sizes in Fig. \ref{hist18cm}. The distribution of the spectral index of the known SNR and H{\sc ii} regions within M82 as a function of brightness temperature was studied by \cite{mcdonald02}, who showed the distribution to be distinctly bi-modal and hence the study of the spectra can be used as an indication of the nature of the source. The unknown sources from this dataset are currently lacking sufficient spectral information to distinguish them fully as either {\HII} regions or SNR. The source 45.39+60.3 does have a three point spectral energy distribution published in \cite{allen99} suggesting it has a flat spectrum and is therefore an {\HII} region, though the brightness temperature measured here ($\sim 5\times10^{4}$\,K) is larger than expected for an {\HII} region and thus there classification remains ambiguous. \cite{allen99} also detected the source 39.47+55.4 at 3.6\,cm and 6\,cm, which combined with the flux density of 0.56$\pm$0.13\,mJy presented here would suggest a flat spectrum for this source. The flux density and size measured implies a brightness temperature of $\sim 10^{4}$\,K, consistent with that of an {\HII} region.  The remaining unknown source is detected for the first time in these observations and shows what could be interpreted as knots of radio emission within a shell structure, making it a strong SNR candidate. This is a faint source with a large ($\sim$5\,pc) diameter and a steep spectrum expected of a SNR would make it difficult to detect in higher frequency observations, providing a possible explanation for its prior non-detection.

\section{The compact sources}\label{compact}

The more compact sources within M82 will now be discussed in view of their observed properties at milli-arcsecond resolutions using the global VLBI and previous EVN observations

\subsection{41.95+57.5}

This source is the most compact and until recently the brightest source within M82. It was first discovered in 1968 \citep{bash68}. The EVN observations from 1986 and 1997 do not completely resolve the elongated structure of this source and its unusual morphology is only clearly seen in the higher resolution global VLBI epochs. This source is somewhat atypical and does not possess the shell morphology seen in other supernova remnants, showing a distinctly double-lobed structure. In addition, it has a measured expansion velocity of 1500$\pm$400\,\kms \citep{beswick06}, lower than typically expected for a SNR. It is possible that combined with the large and persistent flux density decay of $\sim$ 8.5\%$\rm{yr^{-1}}$, that this source may represent the remnant of a gamma ray burst \citep{muxlow05,fenech07} . The full discussion of the nature of this source is deferred to a later paper.

\subsection{43.18+58.3}

This source, along with 44.01+59.6 and 45.17+61.2, was only marginally detected in the first epoch of EVN observations and by 1997 showed three weak peaks of emission in 15\,mas resolution images \citep{pedlar99}. There has been no subsequent detection of this SNR in these very high resolution studies. However, this source can be clearly seen in the combined 1.7\,GHz observations (see Fig. \ref{conts}), which reveals a distinct shell-like structure with three peaks of emission and a radius of $\sim$53\,mas. Comparison of two epochs of MERLIN 5\,GHz observations have shown this source to be expanding at 10,500$\pm$3000\,\kms \citep{fenech08}.

\subsection{43.31+59.2}

\begin{figure}
\begin{center}
\includegraphics[width=5.0cm]{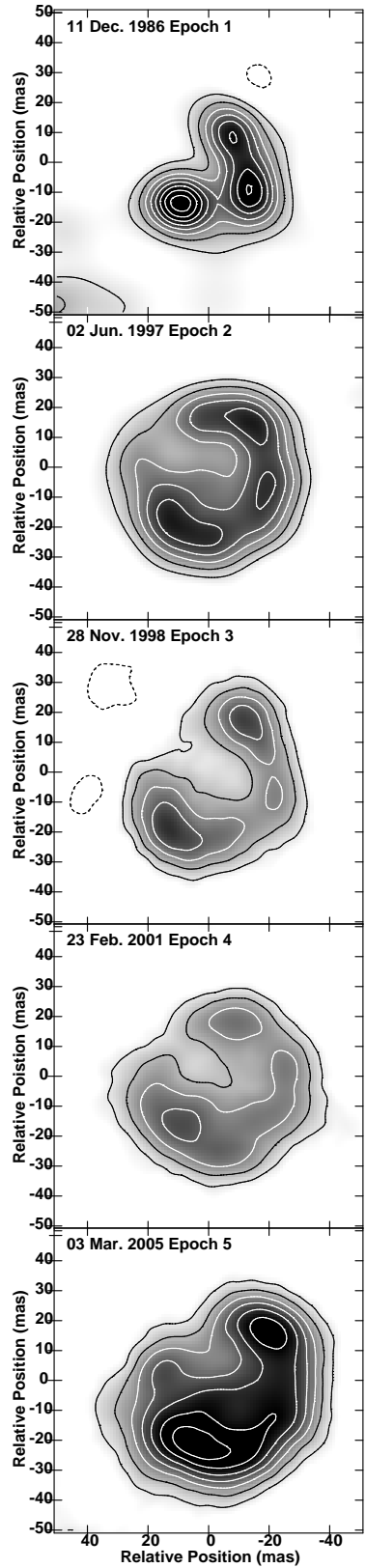}
\caption[Contour and grey-scale images of the SNR 43.31+59.2 from all five epochs of 1.4\,GHz (first epoch only) and 1.7\,GHz observations.]{Contour and grey-scale images of the SNR 43.31+59.2 from all five epochs of 1.4\,GHz (first epoch only) and 1.7\,GHz observations. Each image is restored with a 15\,mas beam. Contours are plotted at $-1,\,1,\,2,\,3,\,4,\,5,\,6,\,7,\,8,\,9,\,10\,\times\,0.35\,\rm{mJy\,beam^{-1}}$ for the EVN observations and $\times0.2\,\rm{mJy\,beam^{-1}}$. The grey-scale is linear ranging from 0.1 to 2.2\, $\rm{mJybeam^{-1}}$ for the EVN epochs and from 0.1 to 1.2\,$\rm{mJybeam^{-1}}$ for the global VLBI epochs.}
\label{cl15}
\end{center}
\end{figure}

This SNR was first observed in images of M82 in 1972 \citep{kronberg75}, though has remained unresolved in MERLIN and VLA studies of the galaxy. However, the EVN and global VLBI observations discussed here completely resolve this source, showing a well-defined almost complete shell structure (see Fig. \ref{cl15}). This source is an excellent example of a shell-like SNR and provides an ideal opportunity to study the evolution of a relatively young, rapidly evolving remnant. This will be discussed in detail in the following section.

\subsubsection{The expansion of 43.31+59.2}\label{expsec}

\begin{figure}
\begin{center}
\includegraphics[width=5cm,angle=270]{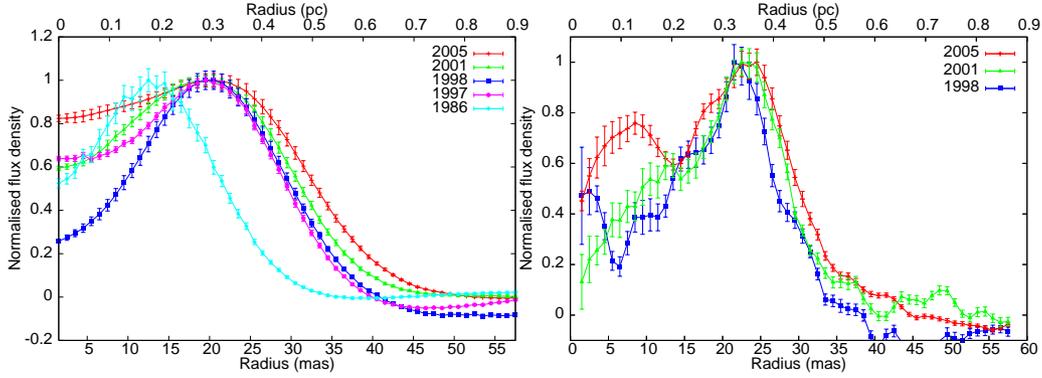}
\caption[Azimuthally-averaged radial profiles of 43.31+59.2 at all epochs from 15\,mas resolution images as well as from the global VLBI epochs at 4\,mas resolution.]{Azimuthally-averaged radial profiles of 43.31+59.2 at all epochs from 15\,mas resolution images as well as from the global VLBI epochs at 4\,mas resolution. The error-bars are 1${\rm{\sigma}}$ errors on the measured flux density around the annulus at a specific radius as derived by {\sc{IRING}} for each 1\,mas increment in radius It should be noted that these points are not independent as their separation is less than the beamsize.}
\label{iring15}
\end{center}
\end{figure}

\begin{figure}
\begin{center}
\includegraphics[width=5.5cm,angle=90]{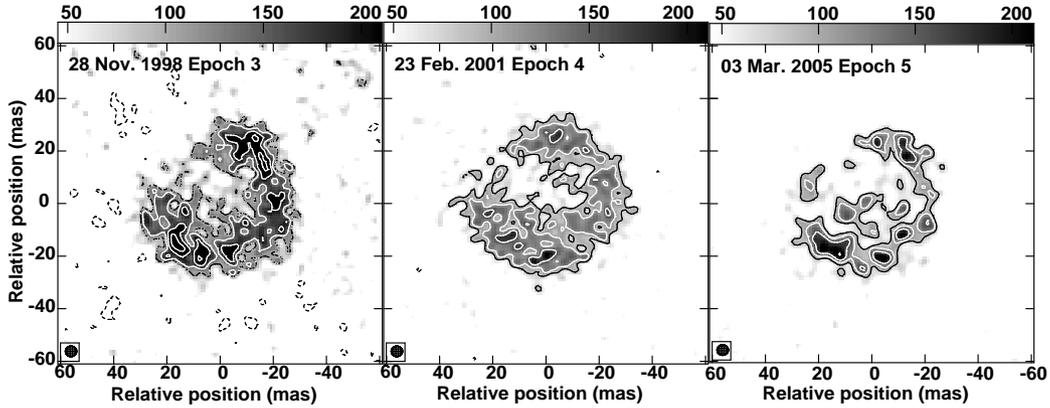}
\caption[Contour and grey-scale images of the SNR 43.31+59.2 from the global VLBI epochs, restored with a 4\,mas beam.]{Contour and grey-scale images of the SNR 43.31+59.2 from the global VLBI epochs, restored with a 4\,mas beam. The images have been contoured with multiples of $-1,\,1,\,1.414,\,2,\,2.828,\,4$ and $5.657 \times\,0.1\,\rm{mJy\,beam^{-1}}$ for the 1998 epoch and $\times\,0.08\,\rm{mJy\,beam^{-1}}$ for the other two epochs.}
\label{43-4}
\end{center}
\end{figure}

In order to determine the size and therefore expansion of 43.31+59.2 the {\sc {aips}} task {\sc{iring}} was used to measure radial profiles (averaged in azimuth) of the source at each of the five epochs using the 15\,mas resolution images. The centre of the source was determined as the point equidistant from the positions of the peaks of radio emission in each image, which were measured using Gaussian fitting. Radial profiles were then measured from each pixel in 1\,mas separations, in a 9$\times$9 grid centred on this point. As discussed in detail in \cite{beswick06}, this process was performed in order to assess the robustness of the use of the radial profiles to measure the shell radius when offset from the centre. 
The radial flux density profiles produced from the annuli at the central point are shown in Fig. \ref{iring15} and the sizes measured from the peak and 50\% of the peak, in each case, are listed in Table \ref{highsize}. This process was also performed using the higher resolution 4\,mas images for only the global VLBI epochs, shown in Fig. \ref{43-4}. The errors listed in Table \ref{highsize} represent the 1$\rm{\sigma}$ deviation of the sizes as measured from all of the radial profiles in the grid. As can be seen in Table \ref{highsize}, measured values of the radius-at-peak of these profiles is systematically slightly larger when measured from the 4\,mas resolution image than the 15\,mas image. This effect results from the smoothing of the data and hence the increased contribution of fainter more diffuse emission filling the SNR shell \citep[see Fig. 7 of ][]{beswick06}. This effect is minimised if the size at 50\% of the peak is used. It should be noted that this has no systematic affect on the expansion velocities found using similarly derived size measurements between epochs.
The evolution of the small-scale structure visible at full global VLBI resolutions makes identifying common knots of radio emission between epochs difficult. As a consequence, it has not been possible to perform Gaussian fitting to the four knots of radio emission initially identified by \cite{mcdonald01} to follow their expansion. 

\begin{tabfonta}
\begin{table}
\begin{center}
\caption[Measured radii of the SNR 43.31+59.2 from the 15\,mas and 4\,mas resolution images, derived using the integrated radial profiles.]{Measured radii of the SNR 43.31+59.2 from the 15\,mas and 4\,mas resolution images, derived using the integrated radial profiles shown in Fig. \ref{iring15}.}
\begin{tabular}{|c|c|c|c|c|c|c|}
\hline
\multicolumn{1}{|c}{Epoch} &
\multicolumn{2}{|c}{4\,mas resolution} &
\multicolumn{4}{|c}{15\,mas resolution} \\

\multicolumn{1}{|c}{} &
\multicolumn{2}{|c}{Radius at peak} &
\multicolumn{2}{|c}{Radius at peak} &
\multicolumn{2}{|c|}{Radius at 50\% of peak}\\

 & (mas) & (pc) & (mas) & (pc) & (mas) & (pc) \\
\hline
1 & -- & -- & 13.2\,$\pm$\,1.0 & 0.205\,$\pm$\,0.016 & 22.0\,$\pm$\,0.9 & 0.343\,$\pm$\,0.014 \\  
2 & -- & -- & 19.2\,$\pm$\,0.6 & 0.300\,$\pm$\,0.009 & 29.5\,$\pm$\,0.8 & 0.457\,$\pm$\,0.012 \\
3 & 21.5\,$\pm$\,0.3 & 0.333\,$\pm$\,0.005 & 19.5\,$\pm$\,0.9 & 0.302\,$\pm$\,0.014 & 30.1\,$\pm$\,0.9 & 0.467\,$\pm$\,0.014 \\
4 & 22.5\,$\pm$\,0.3 & 0.349\,$\pm$\,0.005 & 20.0\,$\pm$\,0.8 & 0.310\,$\pm$\,0.012 & 31.4\,$\pm$\,0.8 & 0.487\,$\pm$\,0.012\\
5 & 24.5\,$\pm$\,0.4 & 0.380\,$\pm$\,0.006 & 20.8\,$\pm$\,0.9 & 0.322\,$\pm$\,0.014 & 33.5\,$\pm$\,0.9 & 0.519\,$\pm$\,0.014 \\
\hline
\end{tabular}
\label{highsize}
\end{center}
\end{table}
\end{tabfonta}

As can be seen in the contour images and from the information in Table \ref{highsize}, the radio shell of 43.31+59.2 visibly expands between the first and subsequent epochs. The azimuthally-averaged radial profiles from the higher angular resolution (4\,mas) images have increased in peak radius from 22.5\,mas to 24.5\,mas between 2001 and 2005, corresponding to an angular expansion rate of 0.50$\pm$0.12\,mas\,$\rm{yr^{-1}}$. This results in an expansion velocity of 7600$\pm$1800\,\kms. Determining the expansion from the peak of the lower resolution (15\,mas) radial profiles is more difficult but using the measurements at 50\% of the peak, also shows a marginally significant expansion for the latest two epochs, consistent with measurements made on the higher (4\,mas) resolution images. This is in good agreement with previous measurements from EVN \citep[e.g. 9850$\pm$1500\,\kms from ][]{pedlar99} and global VLBI epochs, as well as the measurement from MERLIN 5\,GHz observations \citep{fenech08}. 

\subsubsection{Possible deceleration}\label{dec}

A supernova remnant is expected to follow a size evolution described by \citep{huang94} \begin{equation}D=kt^{m}\end{equation} where $D$ is the shell diameter in pc, $k$ is a constant, $t$ is the age in years and $m$ is the deceleration parameter. Free expansion of a supernova remnant is described by $m=1$ and values of $m<1$ describe various stages of the evolution of a SNR as the radio shell begins to decelerate as a result of strong interaction with the material surrounding the SNR. When the SNR enters the Sedov phase of the evolution, the expected deceleration parameter is $\sim$0.45. 

\begin{figure}
\begin{center}
\includegraphics[width=8cm,angle=270]{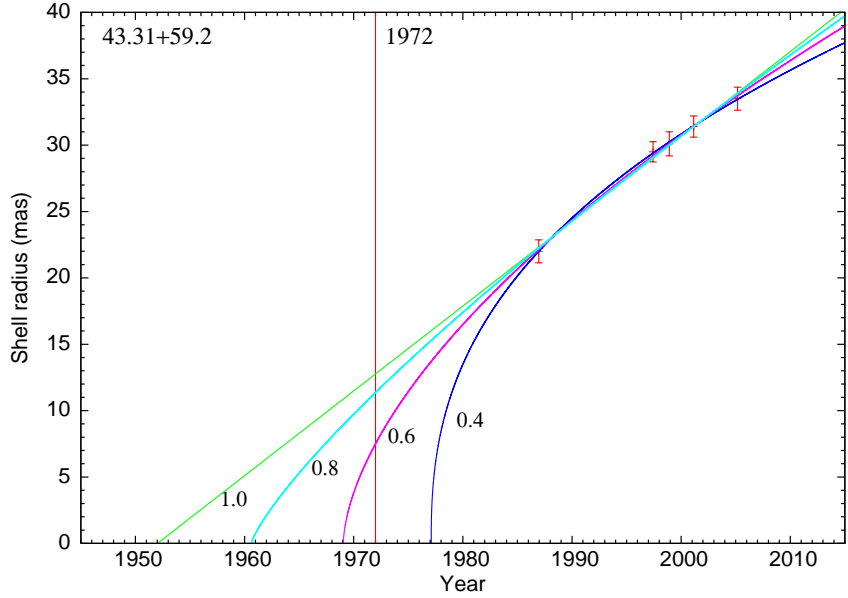}
\caption[The expansion of the source 43.31+59.2, showing the measured sizes from each epoch and the deceleration parameter fits to the data.]{The expansion of the source 43.31+59.2, showing the measured sizes from each epoch and the deceleration parameter fits to the data.}
\label{decp}
\end{center}
\end{figure}

The size measurements for the SNR shell of 43.31+59.2 can be used to determine if it is still freely expanding or has begun to decelerate. Fig. \ref{decp} shows the sizes from Table \ref{highsize} and a number of possible deceleration parameters fitted to the data. The SNR was first detected in observations of M82 made in 1972 \citep{kronberg75}, providing a lower limit to the age of this source. Incorporating this source age limit with the fit to the observed shell size provides a lower limit to the deceleration parameter of 0.53$\pm$0.06. As can be seen from Fig. \ref{decp} it is still very difficult to distinguish between the possible values and hence whether the SNR has actually begun to decelerate. Therefore, assuming the source to be in free expansion gives an approximate birth date of $\sim$\,1952, giving an upper limit to the age of 53\,yrs in 2005.

\begin{figure}
\begin{center}
\includegraphics[width=8cm,angle=270]{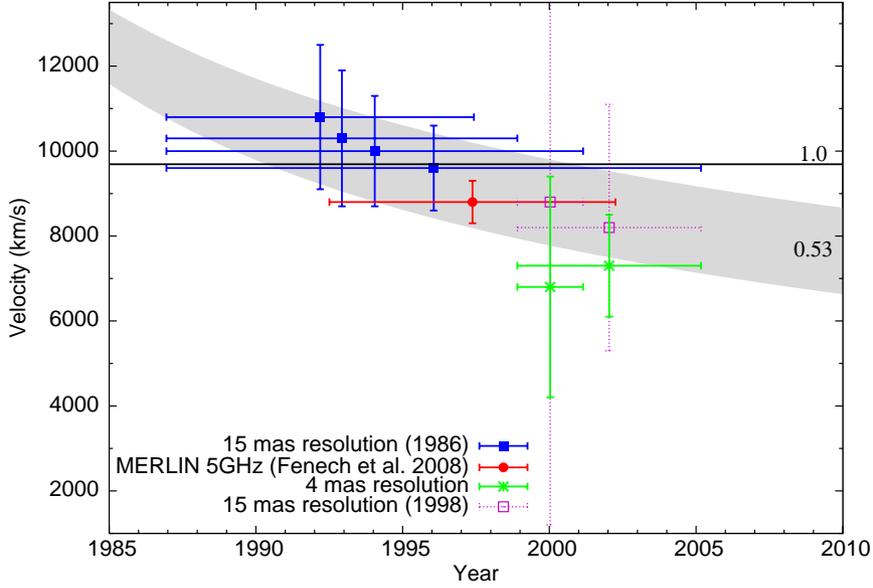}
\caption[Expansion velocities of the SNR 43.31+59.2]{Expansion velocities for the SNR 43.31+59.2. The vertical bar shows the error on the velocity measurement and the horizontal bar represents the timerange between the size measurements used. Velocities from all available epochs have been included as well as the measurement from MERLIN 5\,GHz observations \protect{\citep{fenech08}}. The shaded region represents the trend in velocity determined by a deceleration parameter of 0.53$\pm$0.06. Velocities from 15\,mas measurements between 1986 and each subsequent epoch have been plotted. The 15\,mas equivalent velocities to those from the 4\,mas sizes have also been included for direct comparison, these are from measurements between 1998 and each subsequent epoch.}
\label{vels43}
\end{center}
\end{figure}

Although in agreement with previous measurements, the velocities calculated from the latest epochs are slightly lower than previously observed and it is therefore necessary to investigate the possibility that the radio-shell has begun to decelerate. Fig. \ref{vels43} shows the expansion velocities measured from the {\sc{iring}} profiles between 1986 and each subsequent epoch at 15\,mas resolution. The velocities measured between each epoch at 4\,mas are also shown. The shaded region represents the expected trend of the velocity from a radio shell determined by the deceleration parameter calculated from the size measurements (including errors). Though inconclusive at this stage, this plot does suggest that the SNR may be decelerating.

\subsection{44.01+59.6}

\begin{figure}
\begin{center}
\includegraphics[width=14.5cm,angle=0]{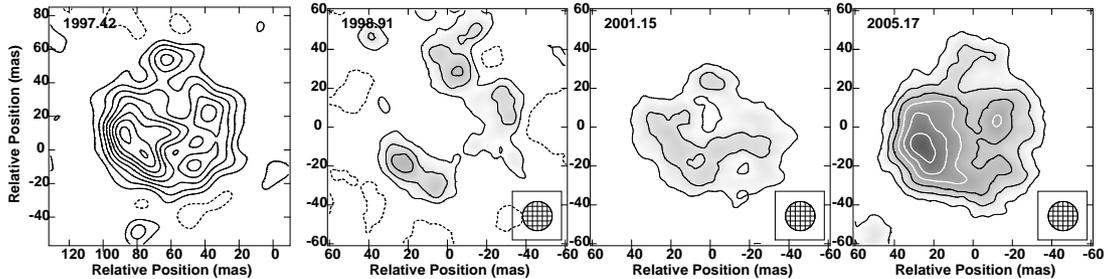}
\caption[Contour and grey-scale images of the source 44.01+59.6 from the EVN (1997) and three global VLBI epochs, restored with a 15\,mas beam.]{Contour and grey-scale images of the source 44.01+59.6 from the three global VLBI epochs, restored with a 15\,mas beam. Contours are plotted at $-1,\,1,\,2,\,3,\,4,\,5,\,6,\,7,\,8,\,9,\,10\times\,0.1\,\rm{mJy\,beam^{-1}}$.}
\label{44.01fig}
\end{center}
\end{figure}

This source was detected in the second epoch of EVN observations from 1997, and further investigated in 1998 and 2001, however the latter detections were close to the noise level and could not be used to perform a reliable test for any expansion between the epochs of observations. Fig. \ref{44.01fig} shows the source 44.01+59.6 from the 1997 EVN observations \citep{pedlar99} and the subsequent global VLBI observations in 1998, 2001 and 2005. The latest global VLBI image in particular shows very similar structure to the EVN image from 1997 and should now be able to be used to investigate any expansion between these two epochs. A relatively low expansion velocity of 2700$\pm$650\,\kms was measured for this source using MERLIN 5\,GHz observations. The only lower measured velocity for a SNR in M82 is that of the peculiar source 41.95+57.5. These and future global VLBI studies of this source will enable a more detailed analysis. 

\subsection{45.17+61.2}

\begin{figure}
\begin{center}
\includegraphics[width=14.6cm,angle=0]{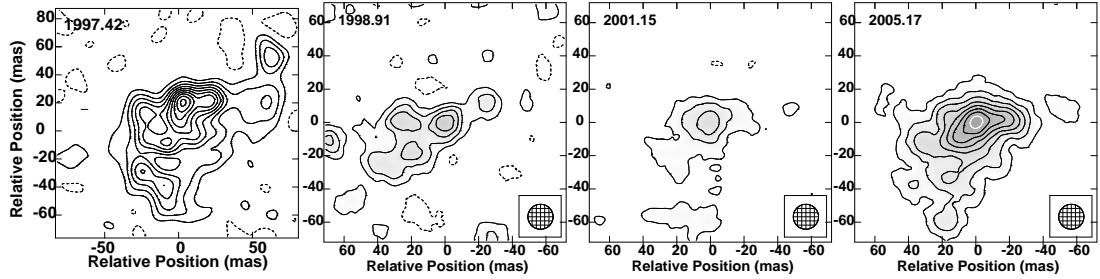}
\caption[Grey-scale and contour images from the global VLBI epochs of 45.17+61.2 restored with a 15\,mas beam.]{Grey-scale and contour image of 45.17+65.3 from the EVN (1997) and global VLBI epochs, restored with an 15\,mas beam.}
\label{45.17fig}
\end{center}
\end{figure}

This source has a more complex structure (see Fig. \ref{45.17fig}) than an obvious shell-like morphology of for example, the SNR 43.31+59.2 (see Fig. \ref{cl15}). The elongation of this source has led to its identification as a possible partial shell of a larger SNR, and assuming this to be the case, monitoring of the position of the peak of emission at 5\,GHz has led to an expansion velocity of $\sim$6000\,\kms \citep[][]{fenech08}. The use of self-calibration in the data reduction degrades the absolute astrometry making calculation of the expansion in this way using the global VLBI data difficult. However, these EVN/global VLBI observations allow a more detailed study of the internal structure of this source, which has clearly evolved between 1997 and 2005, though its overall morphology is still consistent with a partial shell interpretation. 

\section{The properties of the ISM in M82}\label{ism}

\subsection{Ionised gas}\label{ion}

SNR characteristically exhibit a steep spectrum following the power law $S\propto \nu^{\alpha},$ where $\alpha$ is the spectral index. Hence, the SNR previously observed at 5\,GHz \citep{muxlow94,mcdonald02,fenech08} (and those below the detection threshold) are expected to be brighter at 1.7\,GHz. However, not all of the known SNR detected at higher frequencies at similar resolutions have been observed in these data. As the noise levels of the images presented here are greater than for some of the previously published data at higher frequencies, there is an inherent selection effect toward the brighter sources, hence fewer of the fainter SNR and {\HII} regions are detected.

\begin{table}
\begin{center}
\caption[A comparison of the number of sources detected in these 1.7\,GHz and the 5\,GHz MERLIN observations.]{A comparison of the number of sources detected in these 1.7\,GHz and the 5\,GHz MERLIN observations presented in \cite{fenech08}.}
\begin{tabular}{|c|c|c|}
\hline
Source & 1.7\,GHz & 5\,GHz \\
\hline
SNR & 31 & 37 \\
\HII\,\,regions & 2 & 13 \\
Unknown & 3 & 4 \\
\hline
Total & 36 & 55\\
\hline
Shell structures & $\sim$20 & $\sim$28\\
\hline
\end{tabular}
\end{center}
\label{num}
\end{table}

In addition to the known thermal {\HII} regions, extensive regions of molecular and ionised gas have been identified within the central kpc of M82 from the 92\,GHz continuum studies by \cite{carlstrom91}, where the emission is dominated by free-free emission, as well as via radio recombination line (RRL) observations (\citealp*[e.g. the 1.7 and 5\,GHz RRLs in][]{seaquist85}, as well as those presented by \citealp{rodr04}). 
Low frequency observations of M82 \citep{wills97,noglik96}, have shown the presence of significant free-free absorption by ionised gas. The most striking example of this is the large `hole' observed at 408\,MHz and 327\,MHz, roughly centred on the source 41.95+57.5 and approximately 100\,pc in diameter. \cite{wills97} attribute this to free-free absorption from a large {\HII} region ionised by a cluster of early-type stars. 

Analysis of the spectral energy distributions (SEDs) have been performed for a number of the sources within M82 such as those by \cite{wills97,mcdonald01,allen98,allen99} and \cite{tsai09}. These are used to distinguish the flat spectrum {\HII} regions from the steep spectrum SNR. Such studies also include modelling of the free-free absorption seen for individual sources providing an estimate of the amount of ionised gas along the line of sight via the emission measure.

The spectrum of a number of sources show low-frequency turnovers, in the case of SNR suggesting free-free absorption, indicating that some of the SNR are located either within or even behind the ionised gas. A total of nine of the known SNR within M82 show a turnover in their spectrum at frequencies greater than 1.7\,GHz, eight of which have been detected in these observations and have reduced flux densities consistent with this picture. Whilst not detected in these observations, the SNR 43.81+62.8 has been detected at a number of frequencies and was suggested to have a turnover frequency of $\sim$2-3\,GHz by \cite{wills97} via a spectral analysis incorporating observations from 0.408, 5 and 8.3\,GHz  \citep{wills97,kronberg85,huang94}. However, later observations reported in \cite{allen99} at similar frequencies suggest the flux density of this source to have increased. In addition, MERLIN observations at 1.4\,GHz in 1995 \citep*{wills98} give a flux density of 2.1$\pm$0.8\,mJy, compared to an upper limit of 0.2\,mJy\,$\rm{beam^{-1}}$ in these observations. It is therefore difficult to attribute the lack of detection of this SNR to free-free absorption alone, as it may well be a result of its apparent variability.

There are twenty previously identified {\HII} regions in M82, all but two of which have been shown to be or become optically thick above 1.7\,GHz \citep[see][]{wills97,mcdonald01,rodr04}. As a result, only two of the {\HII} regions expected to be in the optically thick regime have been detected in the combined 1.7\,GHz observations, the remainder having predicted flux densities below the image noise level. 
The {\HII} regions 45.63+66.9 and 43.21+61.3 identified by \cite{rodr04}, have a spectral index of $\sim$-0.04 between 8.3 and 43\,GHz. The lack of detection here would suggest that these sources also become optically thick at a frequency greater than 1.7\,GHz. 
The low-resolution image produced from the MERLIN-only data has a 1$\rm{\sigma}$ brightness temperature sensitivity limit of $\sim$1500\,K. There is emission at the 2-3$\rm{\sigma}$ level in this MERLIN-only image for two of the {\HII} regions expected to be optically thick, though they are undetected in the MERLIN+global VLBI images. This would suggest that the low brightness temperature sensitivity of the combined data may also be a contributing factor to the lack of detection of some of the {\HII} regions.

\subsection{Supernova remnants and the star formation rate in M82}

It is possible to use these observations to make an independent estimate of the supernova and star formation rates in M82, enabling a comparative measurement to those performed elsewhere, both from radio and other wavelength observations.

The simplest method of calculating the supernova rate using SNR in M82 is by comparison to the Galactic Remnant Cassiopeia A, which has a known age of $\sim$325 years \citep[age in 2004,][]{fesen06}. Assuming that the observed SNR in M82 which are more luminous than Cassiopeia A are therefore younger, the supernova rate can be calculated as the number of SNR divided by the age of Cassiopeia A. Taking a flux density for Cassiopeia A of 1894\,Jy (extrapolated from the 1\,GHz value using a spectral index of $\alpha=-0.77$, from \cite[observation date 1999][]{green04}), this provides 12 M82 SNR and a resulting supernova rate of $\nu_{SN}=0.04\,\rm{yr}^{-1}$. This is lower than that measured from MERLIN 5\,GHz observations \citep[$\sim$0.07$\,\rm{yr}^{-1}$ from][]{fenech08}, though this is likely to be a result of free-free absorption causing low-frequency turnovers in the SNR spectrum (see section \ref{ion}).

\begin{figure}
\begin{center}
\includegraphics[width=8cm,angle=270]{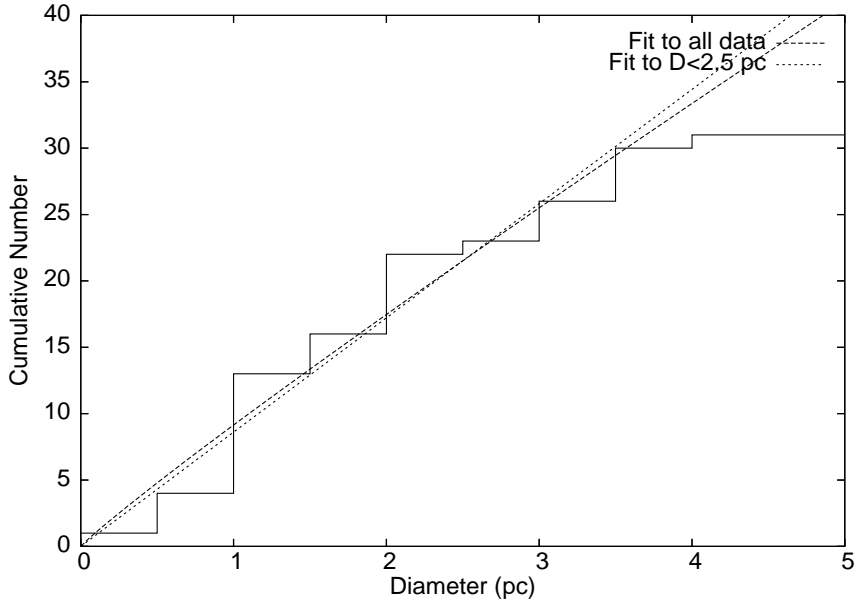}
\caption[The cumulative distribution of SNR at 1.7\,GHz in M82 as a function of diameter.]{The cumulative distribution of SNR at 1.7\,GHz in M82 as a function of diameter, showing the linear and power-law fits to the data for D$<$2.5\,pc.}
\label{cum18cm}
\end{center}
\end{figure}

The supernova rate can also be determined from the cumulative size distribution of the supernova remnants, shown in Fig. \ref{cum18cm}. As discussed in section \ref{dec}, the evolution of the diameter of a SNR with time can be modelled by $D=kt^{m}$, where m is the deceleration parameter and k is a constant. If it is assumed that this applies to all of the SNR in our sample, the cumulative distribution (N($<$D)-D Relation) will follow $N(<D)-D = \nu_{SN}k^{-1/m}D^{1/m}$ where $\nu_{SN}$ is the supernova rate. 
If the SNR are following free expansion, the cumulative distribution can be modelled using a value of $m=1$. This, for the 1.7\,GHz cumulative distribution can be fitted to diameters with D$<$2.5\,pc and provides an estimate of \begin{equation}\nu_{SN}=(0.09\pm0.01)\bigg(\frac{V_{exp}}{5000\,{\rm kms^{-1}}}\bigg)\end{equation} for the supernova rate, where $V_{exp}$ is the expansion velocity. Assuming an expansion velocity of $\sim$5000\,\kms, gives a supernova rate of $\nu_{SN}\sim0.09\,\rm{yr^{-1}}$, in good agreement with those calculated at other radio wavelengths \citep[e.g.][]{muxlow94,fenech08} as well as those predicted by models of the starburst such as \cite{cram98}.

Assuming a Miller-Scalo \citep{millerscalo79} initial-mass function (represented by a power-law with a constant slope given by $-2.3$), these supernova rates can be used to estimate the star-formation rate for M82, giving SFR($M\geq$\solmasit) $\sim$ 1.25-2.25 \solmas$\rm{yr^{-1}}$. This is in good agreement with the $\sim 2$\,\solmas$\rm{yr^{-1}}$ measurements at other radio wavelengths as well as using FIR, UV and H$\alpha$ \citep[for example][]{young96,bell01}.

\subsection{The SNR environment}

The interstellar medium within M82 is believed to be more extreme than in the Milky Way, with pressures of $10^{7}\,\rm{cm^{-3}K}$ observed within the central star-forming region \citep{smith06}. Theoretical studies of M82, such as those of \cite{chevalier01} assume such ISM pressures, along with densities of $\sim 10^{3}\,\rm{cm^{-3}}$ to predict expansion velocities for the observed SNR of $\sim$500\,\kms, which is in stark contrast to those observed, in particular for the SNR 43.31+58.2  \citep[][this paper]{muxlow05,beswick06,fenech08}. 

If the SNR are assumed to be situated within the ionised gas with temperatures of $\sim 10^{4}$K, ambient gas densities of $\sim 10^{3}\,\rm{cm^{-3}}$ are required to provide the level of pressure observed. \cite{chevalier01} alternatively assume that the majority of SNR are evolving into the interclump medium of molecular clouds, citing the association of strong CO emission with the observed radio emission as a good indication of this. However, \cite{weiss01} conclude that most of the CO emission from the central regions of M82 arises from a warm, low density interclump medium with kinetic temperatures of $\sim$150\,K and densities $\sim 10^{3}\,\rm{cm^{-3}}$ in agreement with previous studies such as \cite{mao00}. More recent investigations using formaldehyde line transitions to independently constrain the properties of the dense molecular gas \citep[e.g.][]{muhle07,muhle09} find kinetic gas temperatures of $\sim$200\,K and densities of $\sim 7 \times10^{3}\,\rm{cm^{-3}}$ in strong agreement with the CO studies. Such temperatures and gas densities imply pressures of $\sim 10^{5}-10^{6}\,\rm{cm^{-3}\,K}$, between one and two orders of magnitude lower than those assumed by \cite{chevalier01}.

Following the free expansion phase, a supernova remnant is expected to enter the Sedov phase of its evolution when the mass of swept-up material becomes equivalent to the mass of the ejecta. For an SNR, this is expected to occur whilst interacting with a constant density ISM, at a radius $r_s\simeq 4.1 \times (M_{ej}/n_{0})^{1/3} \rm{pc}$, where $M_{ej}$ is the ejecta mass in units of 10\,\solmas\, and $n_{0}$ is the ISM density in $\rm{cm^{-3}}$. In the case of 43.31+59.2, this stage of evolution has not yet been reached (as shown in section \ref{expsec}), hence an upper limit to the surrounding density can be set of $\leq\rm{250\,cm^{-3}}$, for an ejecta mass of 5\,\solmas. This would suggest that for the required pressures, 43.31+59.2 is not likely to be situated within the ionised gas, a conclusion supported by the lack of low frequency turnover observed in its spectra \citep{wills97,tsai09}.

\subsubsection{A wind-blown bubble?}

It has previously been suggested that the low density environment that 43.31+59.2 appears to be experiencing, may be a result of a wind-blown bubble from the mass-loss of the progenitor star \citep{beswick06}.

The SNR in M82 are expected to be the result of core-collapse supernova events of massive progenitors (typically $>$ 8\solmas). Most stars with masses $>$ 11\solmas will become red supergiants (RSG) in their final stages of evolution and are believed to be the progenitors of the majority of core-collapse supernovae. Such stars (typically OB stars), undergo mass-loss during their main-sequence phase with expected terminal velocities of $\lesssim$ 3000\,\kms and rates of $10^{-6}-10^{-5}$ (\citealp*[e.g.][]{vink00}; \citealp{muller08}).
The subsequent mass-loss during the red-supergiant phase is expected to be slower and more dense, with typical mass-loss rates of $\sim10^{-5}-10^{-4}$\solmas$\rm{yr^{-1}}$ and wind velocities of $<$100\,\kms \citep[e.g.][]{schaller92,dwarkadas05}. Hence prior to the eventual supernova explosion, there will be a complex circumstellar environment consisting of stellar-wind material from various evolutionary stages as well as potentially a collapsed {\HII} region.

Modelling of the pre-supernova circumstellar environment for red-supergiants indicates that there will be a region of RSG wind close to the progenitor with a density profile given by $\rho_{RSG}\sim 5\times 10^{-20} \dot{M}_{-4} r_{17}^{-2} \nu_{1}^{-1}$, reaching distances of a few parsecs \citep*{vanmarle04,dwarkadas05,perez09}. This will be followed by a lower, roughly constant density region, created by the fast main-sequence wind driving a shocked shell into the surrounding medium. This is the wind-blown bubble and reaches distances of a few tens of parsecs from the progenitor star \citep[see e.g.][and references therein]{vanmarle04,vanmarle06,dwarkadas05,perez09}.

Given the sizes of the respective regions and the measured radius of $<$ 1\,pc for 43.31+59.2, this would imply that this source would still be expanding into the remaining RSG wind. However the density gradient would produce a variable flux density, which is not observed. \cite{kronberg00} showed the flux density of 43.31+59.2 and 23 other SNR to be stable over a 10 year period.

The majority of these models assume an interstellar environment more typical of the Galaxy than for a starburst such as M82 and it is therefore possible that the increased pressures found in M82 will have a stronger effect on the development of any wind-blown bubble surrounding the progenitor star \citep[e.g.][]{vanmarle06}. The most likely outcome of which would be a reduced propagation of the various stages of mass-loss into any surrounding {\HII} region and the ISM. Hence it is possible that for the case of 43.31+59.2, the expanding remnant has already entered the lower-density environment of a wind-blown bubble, from which it is yet to emerge.

The interstellar medium within M82 is clearly very complex and whilst this possibility may be argued for a number of other SNR, it may well be a scenario that will not fit all. For example, 44.01+59.6 is comparable in size to 43.31+59.2, though has a measured expansion velocity significantly lower, at $\sim$2700\,\kms \citep{fenech08}, suggesting it's experiencing a higher density environment. A similar point can be made for 41.95+57.5, if it is assumed that it does in fact represent a comparable SNR scenario.

Conversely, sources such as 43.18+58.2 have been shown to be expanding at $\sim$10500\,\kms. This is close to that of the newly discovered radio supernova SN2008iz, which has an average expansion over the first year of $\sim$11000\,\kms (assuming a distance to M82 of 3.6\,Mpc) \citep{brunth09,brunthtel,brunth10}. This implies 43.18+58.2 may still be in free-expansion at a size of $\sim$2\,pc, even though its spectra shows a low-frequency turnover, suggesting it could be situated within or behind the dense ionised gas.

\section{Conclusions}\label{sum}

The relatively young SNR in M82, such as 43.31+59.2, are believed to have ages measured in decades, situating them between the young radio supernovae such as SN\,1979c, and the more evolved SNR within our own Galaxy (e.g. Cass A) which are a few hundred years old. Hence, detailed studies of the SNR in M82 are vital to our understanding of the evolution of SNR in general and in particular, to their early stages of development. The latest global VLBI observations of M82 presented here, have been used to monitor the expansion of the young SNR 43.31+59.2 over a 19 year timeline and show the detailed structure of the sources 44.01+59.6 and 45.17+61.2 and 41.95+57.5.
\begin{enumerate}
\item{The expansion velocity of the SNR 43.31+59.2, has been measured between 2001 and 2005 is 7600$\pm$1800\,\kms. This confirms the high expansion velocity of this SNR measured using previous epochs by \cite{pedlar99,mcdonald01,beswick06}}
\item{Size measurements at each epoch, have been used to study the possible deceleration of 43.31+59.2. Including the 1972 observation of \citep{kronberg75}, a lower limit to the deceleration parameter of 0.53$\pm$0.06 can be fitted to these measurements.}
\item{The SNR 44.01+59.6 has been imaged at each of the global VLBI epochs at 15\,mas resolution. The latest epoch shows a very similar structure to that originally observed by \cite{pedlar99} in 1997.}
\item{The structure of the source 45.17+61.2 has been revealed using the global VLBI 2005 observations confirming its elongated structure.}
\end{enumerate}

In addition the first combined 1.7\,GHz global VLBI and MERLIN observations of M82 have provided detailed images of the individual sources within M82 at this frequency. This has shown many of the SNR to have clear shell or partial shell structures.
\begin{enumerate}
\item{A total of 32 SNR have been detected in these observations with sizes ranging from 0.4 to 4.0\,pc, with a mean diameter of 2.1\,pc as well as two of the known {\HII} regions.}
\item{A possible previously unknown SNR (45.88+69.7) has been detected, with a potential ring-like structure and diameter $\sim$5\,pc.}
\item{The supernova rate has been estimated from these combined observations using several methods which result in measurements of $\nu_{SN}\sim0.04-0.09$. Use of these supernova rates to calculate the SFR for M82 gives values of SFR$(M\geq 5$\,\solmas$)\sim 1.25-2.25\,$\solmas$\rm{yr^{-1}}$ in agreement with measurements from other wavelengths.}
\end{enumerate}

\section*{Acknowledgments}

MERLIN is a national facility operated by The University of Manchester on behalf of the Science and Technology Facilities Council (STFC). The VLBA is operated by the National Radio Astronomy Observatory which is a facility of the National Science Foundation operated under cooperative agreement by Associated Universities Inc. The European VLBI Network is a joint facility of the European, Chinese, South African and other radio astronomy institutes funded by their national research councils. {\sc{parseltongue}} was developed in the context of the ALBUS project, which has benefited from research funding from the European Community's sixth Framework Programme under RadioNet R113CT 2003 5058187. We would like to thank the anonymous referee for the many helpful comments and suggestions that have helped improve this paper.

\bibliographystyle{mn2e}

\label{lastpage}

\end{document}